\documentclass[trackchanges]{aastex701}

\begin{document}

\title{CO Structures with Narrow Lines in Nearby Quiescent Regions}

\author[orcid=0009-0000-2510-2140]{Ruilin Xia}
\affiliation{Purple Mountain Observatory, Chinese Academy of Sciences \\
10 Yuanhua Road, Nanjing 210023, People's Republic of China}
\affiliation{School of Astronomy and Space Science, University of Science and Technology of China \\
96 Jinzhai Road, Hefei 230026, People's Republic of China}
\email[show]{rlxia@pmo.ac.cn}

\author[orcid=0000-0002-0197-470X]{Yang Su}
\affiliation{Purple Mountain Observatory, Chinese Academy of Sciences \\
10 Yuanhua Road, Nanjing 210023, People's Republic of China}
\affiliation{School of Astronomy and Space Science, University of Science and Technology of China \\
96 Jinzhai Road, Hefei 230026, People's Republic of China}
\affiliation{State Key Laboratory of Radio Astronomy and Technology, Purple Mountain Observatory, Chinese Academy of Sciences \\
10 Yuanhua Road, Nanjing 210023, People's Republic of China}
\email[show]{yangsu@pmo.ac.cn}
\correspondingauthor{Yang Su}

\author[orcid=0009-0002-2379-4395]{Shiyu Zhang}
\affiliation{Purple Mountain Observatory, Chinese Academy of Sciences \\
10 Yuanhua Road, Nanjing 210023, People's Republic of China}
\affiliation{State Key Laboratory of Radio Astronomy and Technology, Purple Mountain Observatory, Chinese Academy of Sciences \\
10 Yuanhua Road, Nanjing 210023, People's Republic of China}
\email{syzhang@pmo.ac.cn}

\author[orcid=0000-0003-3151-89645]{Xuepeng Chen}
\affiliation{Purple Mountain Observatory, Chinese Academy of Sciences \\
10 Yuanhua Road, Nanjing 210023, People's Republic of China}
\affiliation{School of Astronomy and Space Science, University of Science and Technology of China \\
96 Jinzhai Road, Hefei 230026, People's Republic of China}
\affiliation{State Key Laboratory of Radio Astronomy and Technology, Purple Mountain Observatory, Chinese Academy of Sciences \\
10 Yuanhua Road, Nanjing 210023, People's Republic of China}
\email{xpchen@pmo.ac.cn}

\author[orcid=0000-0001-7768-7320]{Ji Yang}
\affiliation{Purple Mountain Observatory, Chinese Academy of Sciences \\
10 Yuanhua Road, Nanjing 210023, People's Republic of China}
\affiliation{State Key Laboratory of Radio Astronomy and Technology, Purple Mountain Observatory, Chinese Academy of Sciences \\
10 Yuanhua Road, Nanjing 210023, People's Republic of China}
\email{jiyang@pmo.ac.cn}

\author[orcid=0000-0002-3866-414X]{Yan Gong}
\affiliation{Purple Mountain Observatory, Chinese Academy of Sciences \\
10 Yuanhua Road, Nanjing 210023, People's Republic of China}
\affiliation{State Key Laboratory of Radio Astronomy and Technology, Purple Mountain Observatory, Chinese Academy of Sciences \\
10 Yuanhua Road, Nanjing 210023, People's Republic of China}
\email{ygong@pmo.ac.cn}

\author[orcid=0000-0002-8051-5228]{Yuehui Ma}
\affiliation{Purple Mountain Observatory, Chinese Academy of Sciences \\
10 Yuanhua Road, Nanjing 210023, People's Republic of China}
\affiliation{State Key Laboratory of Radio Astronomy and Technology, Purple Mountain Observatory, Chinese Academy of Sciences \\
10 Yuanhua Road, Nanjing 210023, People's Republic of China}
\email{mayh@pmo.ac.cn}

\author[orcid=0000-0002-3904-1622]{Yan Sun}
\affiliation{Purple Mountain Observatory, Chinese Academy of Sciences \\
10 Yuanhua Road, Nanjing 210023, People's Republic of China}
\affiliation{School of Astronomy and Space Science, University of Science and Technology of China \\
96 Jinzhai Road, Hefei 230026, People's Republic of China}
\affiliation{State Key Laboratory of Radio Astronomy and Technology, Purple Mountain Observatory, Chinese Academy of Sciences \\
10 Yuanhua Road, Nanjing 210023, People's Republic of China}
\email{yansun@pmo.ac.cn}

\author[orcid=0000-0001-8060-1321]{Min Fang}
\affiliation{Purple Mountain Observatory, Chinese Academy of Sciences \\
10 Yuanhua Road, Nanjing 210023, People's Republic of China}
\affiliation{School of Astronomy and Space Science, University of Science and Technology of China \\
96 Jinzhai Road, Hefei 230026, People's Republic of China}
\affiliation{State Key Laboratory of Radio Astronomy and Technology, Purple Mountain Observatory, Chinese Academy of Sciences \\
10 Yuanhua Road, Nanjing 210023, People's Republic of China}
\email{mfang@pmo.ac.cn}

\author[orcid=0000-0002-7489-0179]{Fujun Du}
\affiliation{Purple Mountain Observatory, Chinese Academy of Sciences \\
10 Yuanhua Road, Nanjing 210023, People's Republic of China}
\affiliation{School of Astronomy and Space Science, University of Science and Technology of China \\
96 Jinzhai Road, Hefei 230026, People's Republic of China}
\affiliation{State Key Laboratory of Radio Astronomy and Technology, Purple Mountain Observatory, Chinese Academy of Sciences \\
10 Yuanhua Road, Nanjing 210023, People's Republic of China}
\email{fjdu@pmo.ac.cn}

\author[orcid=0000-0003-2549-7247]{Shaobo Zhang}
\affiliation{Purple Mountain Observatory, Chinese Academy of Sciences \\
10 Yuanhua Road, Nanjing 210023, People's Republic of China}
\affiliation{State Key Laboratory of Radio Astronomy and Technology, Purple Mountain Observatory, Chinese Academy of Sciences \\
10 Yuanhua Road, Nanjing 210023, People's Republic of China}
\email{shbzhang@pmo.ac.cn}

\author[orcid=0000-0003-2418-3350]{Xin Zhou}
\affiliation{Purple Mountain Observatory, Chinese Academy of Sciences \\
10 Yuanhua Road, Nanjing 210023, People's Republic of China}
\affiliation{State Key Laboratory of Radio Astronomy and Technology, Purple Mountain Observatory, Chinese Academy of Sciences \\
10 Yuanhua Road, Nanjing 210023, People's Republic of China}
\email{xinzhou@pmo.ac.cn}

\author[orcid=0000-0003-0804-9055]{Lixia Yuan}
\affiliation{Purple Mountain Observatory, Chinese Academy of Sciences \\
10 Yuanhua Road, Nanjing 210023, People's Republic of China}
\affiliation{State Key Laboratory of Radio Astronomy and Technology, Purple Mountain Observatory, Chinese Academy of Sciences \\
10 Yuanhua Road, Nanjing 210023, People's Republic of China}
\email{lxyuan@pmo.ac.cn}

\author[orcid=0000-0003-4586-7751]{Qing-Zeng Yan}
\affiliation{Purple Mountain Observatory, Chinese Academy of Sciences \\
10 Yuanhua Road, Nanjing 210023, People's Republic of China}
\affiliation{State Key Laboratory of Radio Astronomy and Technology, Purple Mountain Observatory, Chinese Academy of Sciences \\
10 Yuanhua Road, Nanjing 210023, People's Republic of China}
\email{qzyan@pmo.ac.cn}

\author[orcid=0000-0003-2732-0592]{Li Sun}
\affiliation{Purple Mountain Observatory, Chinese Academy of Sciences \\
10 Yuanhua Road, Nanjing 210023, People's Republic of China}
\affiliation{State Key Laboratory of Radio Astronomy and Technology, Purple Mountain Observatory, Chinese Academy of Sciences \\
10 Yuanhua Road, Nanjing 210023, People's Republic of China}
\email{lisun@pmo.ac.cn}

\author[orcid=0009-0000-3311-0159]{Jiancheng Feng}
\affiliation{Shanghai Astronomical Observatory, Chinese Academy of Sciences \\
80 Nandan Road, Shanghai 200030, People's Republic of China}
\email{fengjiancheng@shao.ac.cn}

\begin{abstract}
Using CO data from Phase~I of the Milky Way Imaging Scroll Painting (MWISP) survey, we present a systematic study of molecular structures with narrow lines. 
We identify 57 CO structures, most of which exhibit low densities and subsonic/transonic turbulence.
Among them, structures with large projected areas and diffuse, sheet-like geometries are identified as veil clouds.
The low LSR velocities and the concentration of these CO structures toward both the Galactic center (e.g., Ophiuchus, Aquila) and anticenter (e.g., Cepheus, Taurus) regions suggest a local origin for the sample, as supported by distance measurements of about 200--300\,pc for a subset with relatively large angular extents.
These nearby structures likely arise from large-scale compression driven by past supernova activity within the Local Bubble.
The observed low-velocity-dispersion emission may trace quiescent regions where turbulence has decayed due to a lack of sustained energy injection.
For diffuse veil clouds with an assumed magnetic field of $\sim 10\,\mu\mathrm{G}$, ion--neutral friction may provide an additional mechanism for turbulent dissipation on sub-parsec scales corresponding to their thickness of 0.1--0.3\,pc.
Tracing the atomic-to-molecular transition, veil clouds provide a unique window into the diffuse, quiescent precursor state of dense gas.
They likely represent a widespread but previously overlooked component of the Galactic molecular gas reservoir, with significant implications for cloud formation and evolution, the total mass budget and spatial distribution of molecular gas, and the initial conditions of star formation as a related consequence.

\end{abstract}


\section{introduction}\label{sec:intro}
Molecular clouds~(MCs) play a pivotal role in the interstellar matter cycle~\citep{2011piim.book.....D,2020SSRv..216...50C,2020MNRAS.493.2872C}.
Since its first detection in the 1970s~\citep{1970ApJ...161L..43W}, CO has served as the primary tracer of molecular gas, rather than H$_2$, which is difficult to observe directly from the ground.
Extensive CO surveys now provide a detailed view of the distribution and properties of molecular gas throughout the Galaxy, and have become essential for understanding the link between MCs and star formation~\citep{1987ARA&A..25...23S,1999ARA&A..37..311E,2012ARA&A..50..531K,2015ARA&A..53..583H,2017ApJ...834...57M}.

Observations show that most CO emission lines exhibit supersonic linewidths, which are generally interpreted as evidence of supersonic turbulence~\citep{1974ARA&A..12..279Z}.
The observed velocity dispersions follow a power-law scaling with size, which is generally consistent with a Kolmogorov-type turbulent cascade~\citep{1981MNRAS.194..809L}. 
These observational facts constitute the empirical foundation of turbulence theory for gas clouds~\citep{2004RvMP...76..125M,2004ARA&A..42..211E}.
Turbulence is important in supporting MCs against gravitational collapse, shaping their morphology, driving chemical evolution, and regulating star formation in dense gas~\citep{1992A&A...257..715F,2007ARA&A..45..565M,2012ApJ...761..156F,2016SAAS...43...85K}.

Turbulence in MCs is primarily driven by the conversion of gravitational potential and rotational energy through gas accretion, large-scale interstellar flows, and galactic shear~\citep{2010A&A...520A..17K,2011ApJ...738..101G,2016MNRAS.458.1671K}. 
Additional contributions come from thermal and magnetohydrodynamic instabilities, as well as stellar feedback including protostellar outflows, stellar winds, and supernova shocks~\citep{2014prpl.conf..451F,2015MNRAS.450..504M,2016ApJ...822...11P}.
Supersonic turbulence cascades from large to small scales via shocks and nonlinear interactions. 
At the dissipation scale, turbulent energy converts to heat through viscous and magnetic dissipation and is rapidly radiated away~\citep{1998A&A...340..241J,1999ApJ...524..169M,2009A&A...495..847G,2013A&A...550A.106L,2022A&A...664A.193R}.
Unless continuously driven, supersonic turbulence within MCs dissipates rapidly, settling into a subsonic state that can be identified in observations.

Previous studies of subsonic/transonic turbulence in MCs have primarily targeted relatively dense filaments and cores~\citep{2011A&A...533A..34H,2015A&A...574A.104T,2016A&A...587A..97H,2018A&A...620A..62G,2020ApJ...896..110L,2022ApJ...926..165L,2022A&A...663A..82G}.
And a few narrow-line CO structures have been identified in single-point/mapping observations toward PGCCs (Planck Catalogue of Galactic Cold Clumps;~\citealt{2011A&A...536A..23P,2012ApJ...756...76W,2016A&A...594A..28P,2016ApJS..224...43Z,2018ApJS..234...28L}).
These dense structures are typically embedded within larger, more diffuse molecular gas reservoirs, following the hierarchical nesting of cores within clumps within MCs~\citep{2015ARA&A..53..583H}.
Recent CO surveys, however, have uncovered a distinct population of clouds characterized by unusually narrow lines and diffuse, extended structures(e.g., wispy clouds in~\citealt{2021MNRAS.500.3064S} and curtain MCs in~\citealt{2025AJ....170..239Z}).
The non-thermal linewidths are comparable to or even smaller than the thermal width, implying that these clouds are characterized by subsonic or transonic turbulence.
These structures were missed by earlier surveys due to their limited spectral resolution ($\sim 1~\mathrm{km\,s^{-1}}$) and coarse angular sampling (beam sizes of several arcminutes), leaving their distribution and physical properties largely unconstrained.

In this paper, we present a study of CO structures with narrow lines using data from the Milky Way Imaging Scroll Painting survey~(MWISP;~\citealt{2019ApJS..240....9S,2026ApJS..282...65Y}).
Thanks to its large-scale coverage, high spatial and spectral resolution, and multiple CO isotopologues, we are able to characterize the spatial distribution and physical properties of these clouds, investigate the origin of their subsonic or transonic turbulence, and explore their evolutionary pathways.
Section~\ref{sec:CO data and preprocessing} outlines the MWISP CO datasets and preprocessing steps. 
Section~\ref{sec:results} presents the sample of narrow-line CO structures and analyzes their spatial distribution and physical properties.
In Section~\ref{sec:discussion}, we discuss their potential associations with local large-scale structures and explain the nature of their subsonic or transonic turbulence by constructing a schematic diagram. 
Finally, Section~\ref{sec:summary} summarizes the main conclusions.

\section{CO data and preprocessing}\label{sec:CO data and preprocessing}
Using the PMO 13.7-m millimeter telescope, the MWISP project produced maps of $^{12}$CO ($J=1$--$0$) and its isotopologues over $2300\,\mathrm{deg}^2$ of the Galactic plane ($9.75^\circ \leq l \leq 229.75^\circ$, $|b| \leq 5.15^\circ$) at a moderate angular resolution ($50''$) between 2011 and 2022.
The data are structured in a three-dimensional ($l$-$b$-$v$) cube with grid sizes of $30''$ in $l$ and $b$ and $0.16\,\mathrm{km\,s}^{-1}$ in velocity. 
The typical rms noise levels are $0.47\,\mathrm{K}$, $0.25\,\mathrm{K}$, and $0.25\,\mathrm{K}$ for $^{12}$CO, $^{13}$CO, and C$^{18}$O($J = 1$--$0$), respectively. 
The MWISP I survey data have now been made publicly available~\citep{2026ApJS..282...65Y}.

MC samples are commonly constructed using Gaussian decomposition and hierarchical cluster identification \citep{2017ApJ...834...57M,2019MNRAS.485.2457H,2024AJ....167..220Z}.
We applied the GaussPy+ algorithm \citep{2019A&A...628A..78R,2020A&A...633A..14R} to perform Gaussian decomposition of the velocity components for each pixel in the $^{12}$CO data cube. 
Given the substantial computational cost of performing Gaussian decomposition and clustering on the large-scale dataset, we restrict the velocity range of the data to $|V_{\rm LSR}| \leq 30\,\mathrm{km\,s}^{-1}$.
Here, we selected several sub-cubes exhibiting emission with narrow lines toward the inner Galaxy, covering a total area of approximately $30\,\mathrm{deg}^2$. 
From these sub-regions, we obtained $\alpha_1$ values ranging from $1.91$ to $2.25$ and $\alpha_2$ values between $3.66$ and $3.95$, where $\alpha_1$ denotes the Gaussian kernel size for primary noise reduction of narrow spectral features and $\alpha_2$ represents the secondary smoothing parameter optimized for broad components. 
The mean values, $\alpha_1=1.99$ and $\alpha_2=3.76$, were adopted for the Gaussian decomposition of the dataset that we are interested in.
Subsequently, we applied the Agglomerative Clustering for ORganizing Nested Structures (ACORNS) algorithm \citep{2019MNRAS.485.2457H,2008ApJ...679.1338R} with its default parameters to cluster the Gaussian components obtained from GaussPy+ into coherent MC structures.

\section{results}\label{sec:results}
\subsection{A Sample of CO Structures with Narrow Lines} \label{sec:3.1}

For about 100,000 CO structures extracted from the position--position--velocity space, we calculate the unweighted arithmetic mean and standard deviation of the linewidth (i.e., full width at half maximum; FWHM) on a pixel-by-pixel basis for each.
The resulting distributions of these two quantities across all structures are non-normal, peaking at $1.08\,\mathrm{km\,s^{-1}}$ and $0.33\,\mathrm{km\,s^{-1}}$, where about 15\% of the sample lies below both peaks.
Within this subsample, the peak linewidth from pixel-by-pixel statistics is near $ 0.8\,\mathrm{km\,s^{-1}}$, which we adopted as the threshold for narrow-line pixels.
We then define narrow-line regions as contiguous $3 \times 3$ pixel areas with linewidths below this value.
To further refine our sample, we introduced the parameter narrow\_cov, defined as the fraction of pixels within a structures that exhibit narrow lines. 
Applying the criteria $\texttt{narrow\_cov} \geq 50\%$ and a projected area $\geq 56.25\,\mathrm{arcmin^2}$, we identified 57 CO structures with narrow lines as our final sample.
Table~\ref{tab:1}~\footnote{This table lists 3 examples. A complete machine-readable version of the catalog is available online.} provides a summary of the physical parameters for these structures.

\begin{figure}[ht]
    \centering
    \includegraphics[page=1, width=0.9\textwidth]{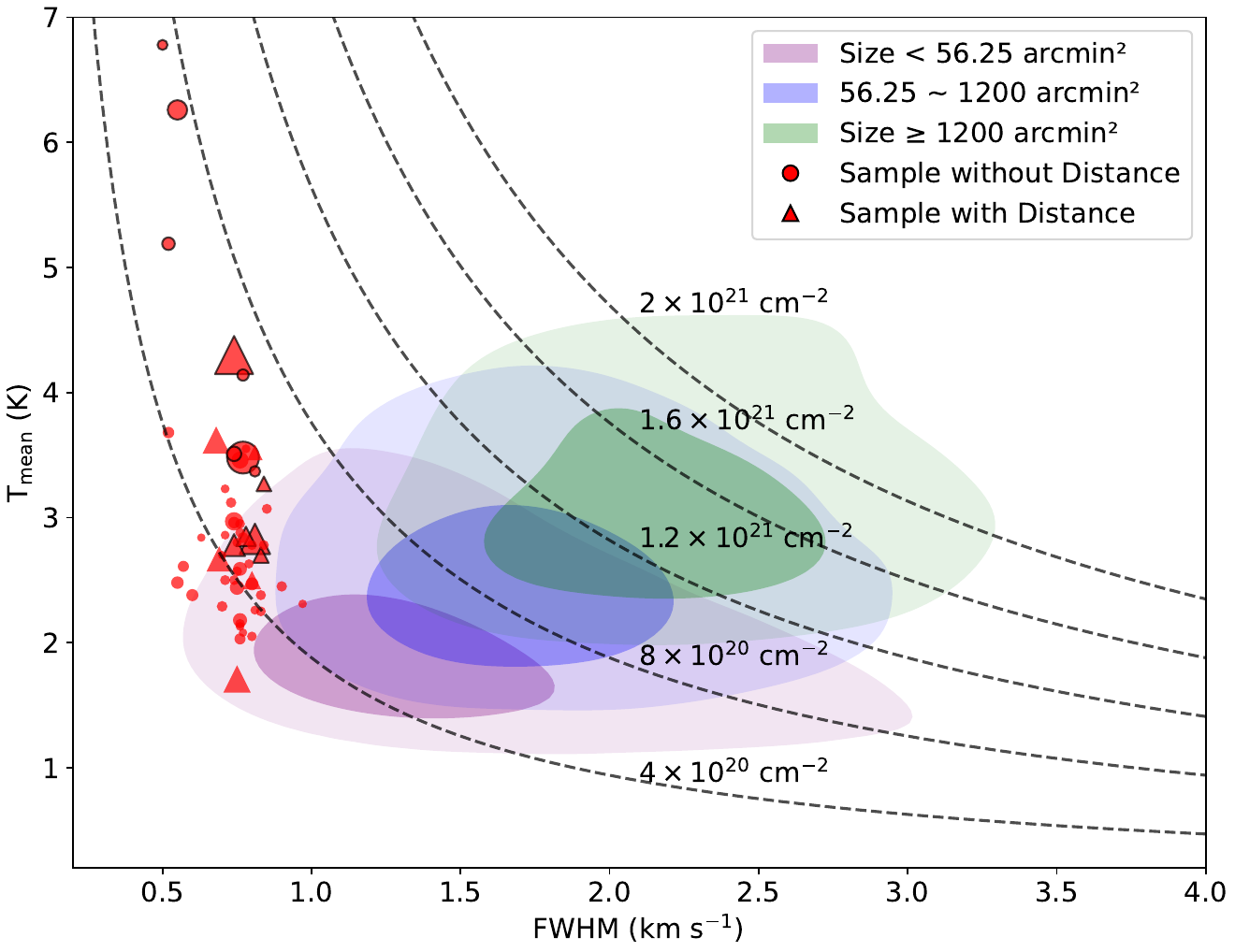}
    \caption{
    Distribution of $^{12}$CO structures in the FWHM~(Column~(6) in Table~\ref{tab:1}) versus mean temperature.
    CO structures with narrow lines are shown as red triangles (with distance) and red circles (without distance), where the symbol size is proportional to the projected area of the $^{12}$CO emission. 
    Clouds outlined in black indicate those with a $^{13}$CO projected area larger than $10~\mathrm{arcmin^2}$. 
    The purple, blue, and green shaded regions mark the kernel density estimation (KDE) of CO
    structures grouped by spatial scale in the FWHM--$T_{\mathrm{mean}}$ plane, where darker shades indicate regions of higher cloud number density (i.e., the densest 30\% according to the KDE). 
    The black dashed line shows the trend of constant H$_2$ column density derived using the $X_{\mathrm{CO}}$-factor (i.e., $X_{\mathrm{CO}} = 2 \times 10^{20}~\mathrm{cm^{-2}\,(K\,km\,s^{-1})^{-1}}$;~\citealt{2013ARA&A..51..207B}). 
    }
    \label{fig:T-FWHM}
\end{figure}

Figure~\ref{fig:T-FWHM} shows the distribution of these CO structures in the FWHM--$T_{\rm mean}$ plane.
Our sample exhibits relatively low column densities and likely lies at the periphery of the main MC distribution (see Section~\ref{sec:3.2}), representing a distinct outlier population.
These structures have $^{12}$CO mean temperatures comparable to those of typical CO structures of similar size, yet their linewidths are even narrower than those measured in smaller-scale CO structures.
Consequently, the H$_2$ column densities inferred from the integrated intensities of these structure (detailed estimation of H$_2$ column density is presented in Section~\ref{sec:3.3.1}) are also low.

\begin{figure}[ht]
    \centering
    {\includegraphics[width=0.3\textwidth]
    {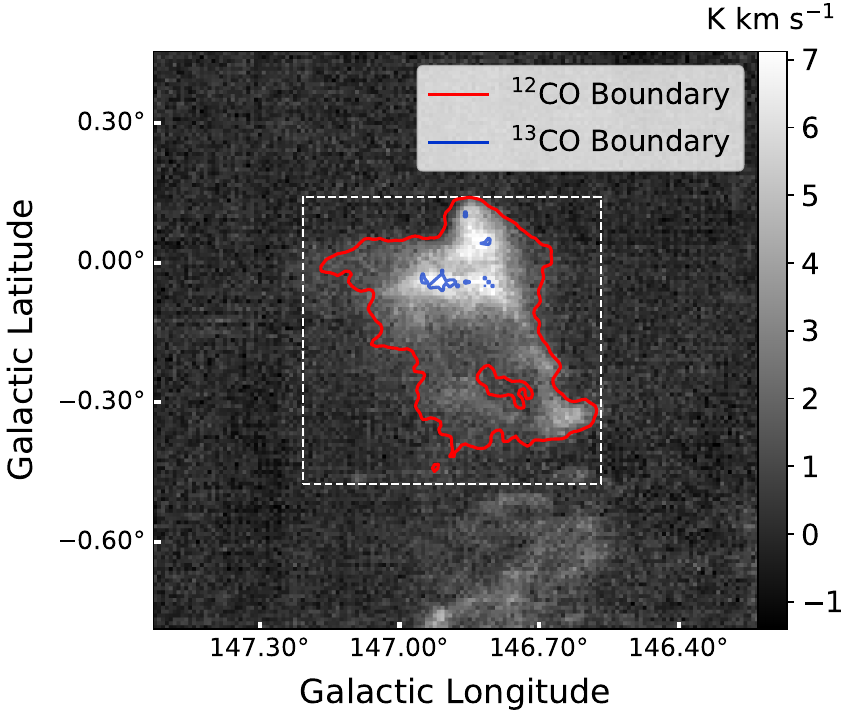}} 
    \quad
    {\includegraphics[width=0.3\textwidth]
    {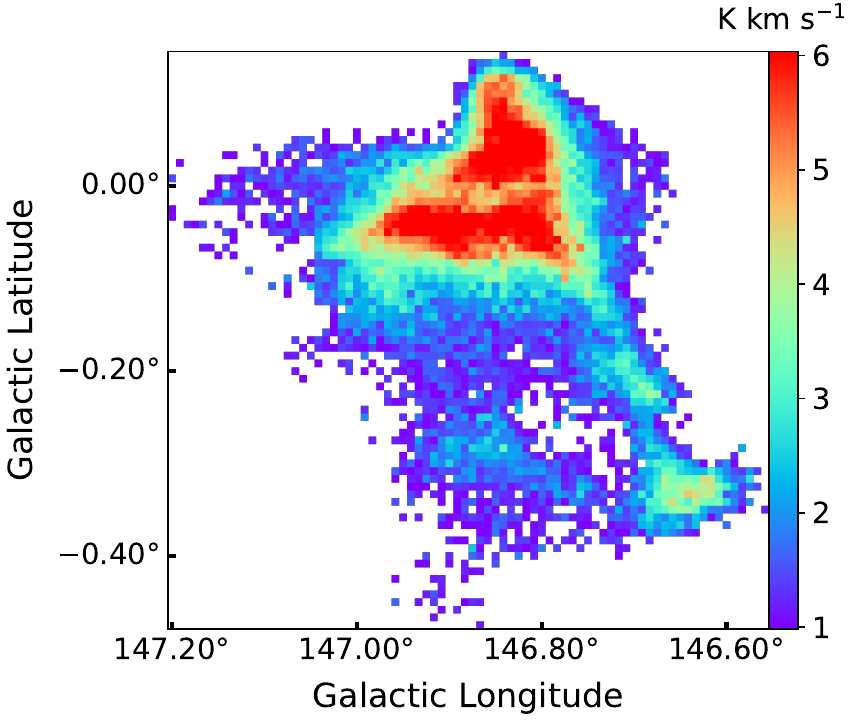}} 
    \quad
    {\includegraphics[width=0.3\textwidth]
    {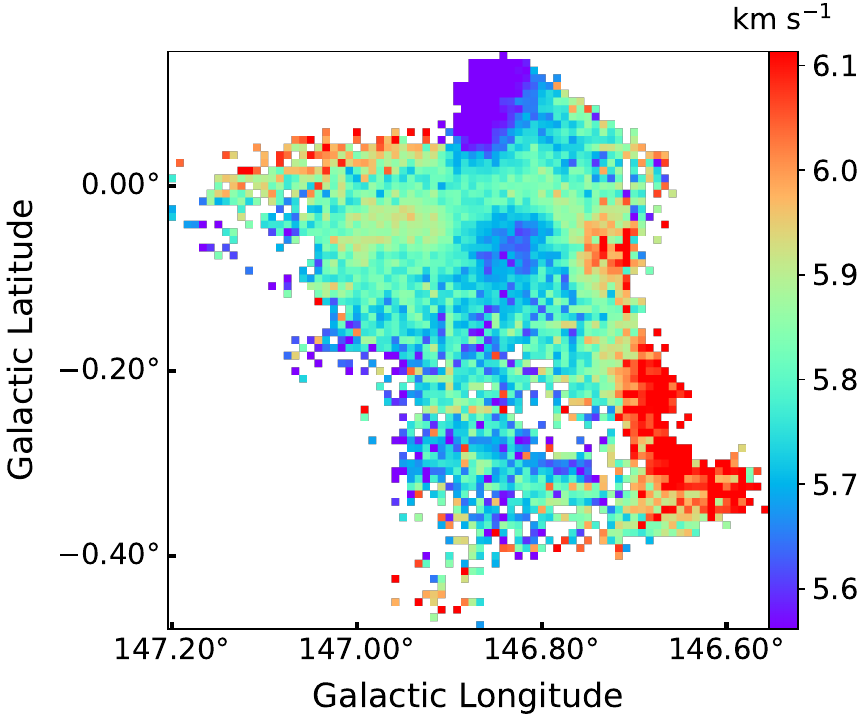}}

    \vspace{0.2cm} 

    {\includegraphics[width=0.3\textwidth]
    {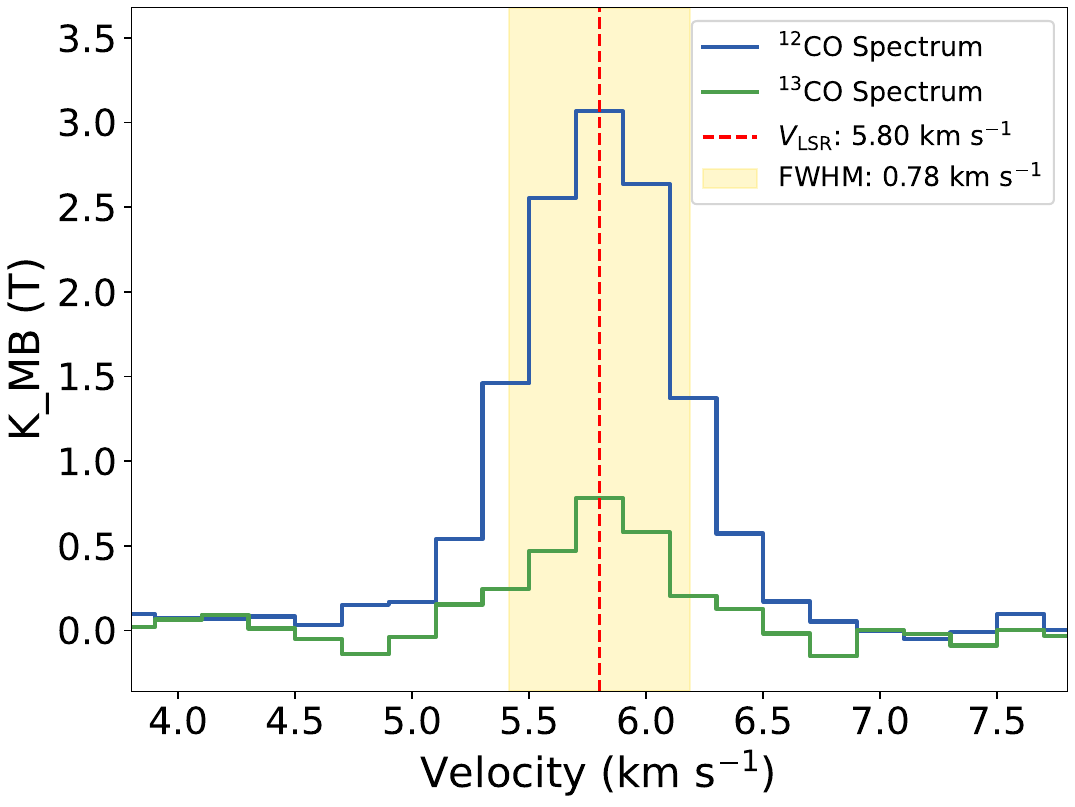}} 
    \quad
    {\includegraphics[width=0.3\textwidth]
    {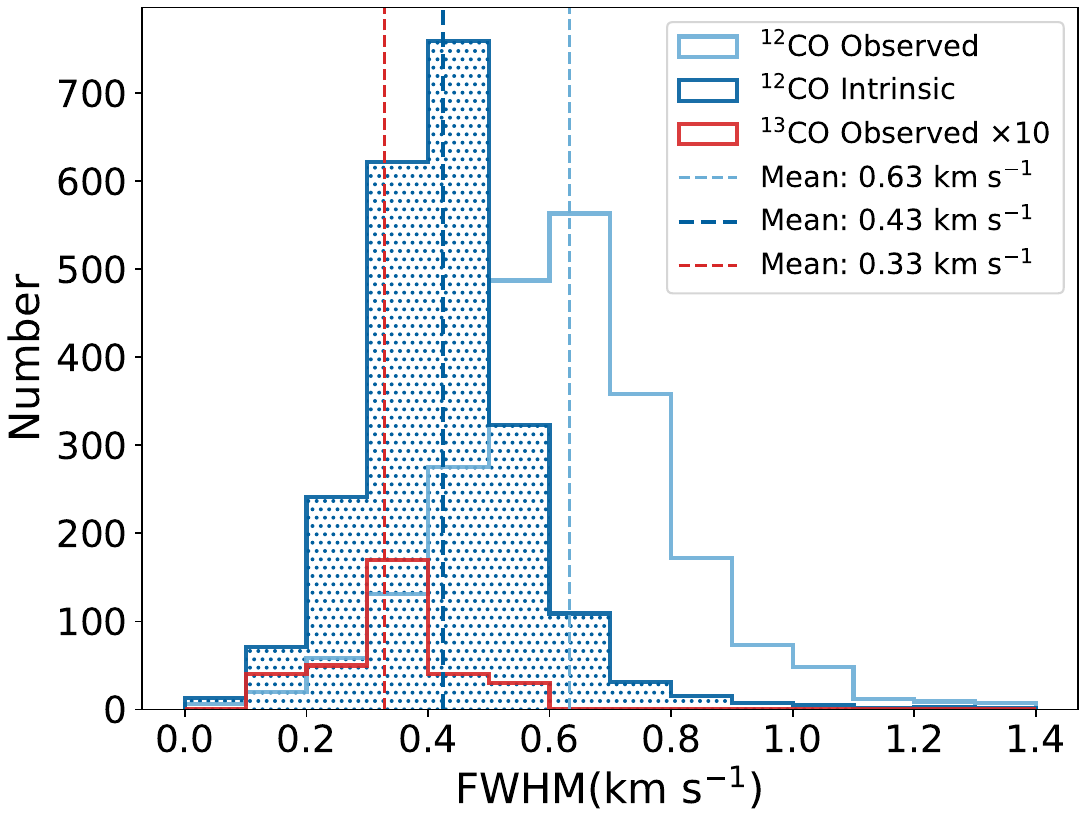}} 
    \quad
    {\includegraphics[width=0.3\textwidth]
    {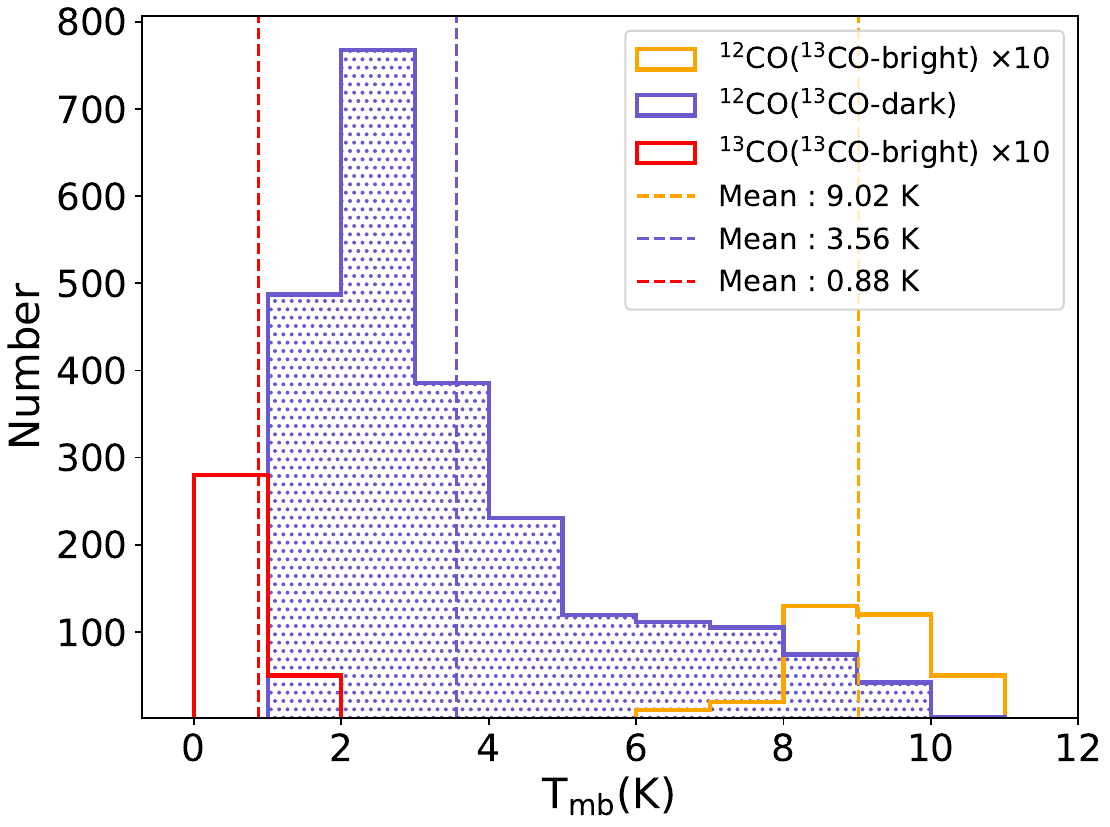}} 

    \caption{
    Top-left: Integrated intensity map of $^{12}$CO emission from cloud G146.845-00.095+005.80~(ID~41) over the velocity interval $4$--$7.5\,\mathrm{km\,s}^{-1}$. 
    The red contour outlines the boundary of the $^{12}$CO structure, while the blue contour shows the boundary of the $^{13}$CO structure extracted within the $^{12}$CO structure.
    Top-middle: Integrated intensity map of $^{12}$CO reconstructed via the GaussPy+ and ACORNS algorithms.
    Top-right: Centroid velocity of $^{12}$CO.
    Bottom-left: Average spectra of $^{12}$CO (blue) and $^{13}$CO (green), smoothed to $0.2\,\mathrm{km\,s}^{-1}$, with a centroid velocity of $5.80\,\mathrm{km\,s}^{-1}$ and an FWHM of $0.78\,\mathrm{km\,s}^{-1}$ (Here, $\mathrm{FWHM} = 2\sqrt{2\ln 2}\,v_{\rm disp}$, derived from the whole cloud velocity dispersion, see Column (5) in Table~\ref{tab:1}).
    Bottom-middle: Pixel-by-pixel linewidth distributions for $^{12}$CO (light blue), $^{13}$CO (red), and opacity-corrected $^{12}$CO in $^{13}$CO-dark regions (dark blue). The number of data points from $^{13}$CO-bright regions in each bin is scaled by a factor of 10 relative to the original count.
    Bottom-right: Brightness temperature distributions of $^{12}$CO in $^{13}$CO-dark region (purple), and $^{12}$CO in $^{13}$CO-bright region (orange), and $^{13}$CO (red).
    }
    \label{fig:410845}
\end{figure}

Specifically, our sample covers an area of over 10,000~$\mathrm{arcmin^2}$, with the largest structures reaching a size of $\sim 1,200\,\mathrm{arcmin^2}$.
Among them, 39 $^{12}$CO structures have detectable $^{13}$CO emission (defined as integrated intensity $> 0.25\,\mathrm{K\,km\,s^{-1}}$), which occupies about 10\% of the area of their $^{12}$CO counterpart. 
Only 13 of the $^{13}$CO structures are larger than 10~$\mathrm{arcmin^2}$, with the largest covering more than 300~$\mathrm{arcmin^2}$.
Figure~\ref{fig:410845} presents a detailed view of a typical region of extended, diffuse $^{12}$CO emission, within which several $^{13}$CO structures are embedded.

As shown in Figure~\ref{fig:fwhm dis}, the $^{12}$CO FWHM (see Column~(6) in Table~\ref{tab:1}) for our sample ranges from 0.50 to 0.97~$\mathrm{km\,s^{-1}}$, peaking at around 0.75~$\mathrm{km\,s^{-1}}$.
In contrast, the $^{13}$CO FWHM for the 13 large-area $^{13}$CO structures is primarily between 0.29 and 0.40~$\mathrm{km\,s^{-1}}$.
A pixel-by-pixel analysis shows that the $^{12}$CO brightness temperature in our sample spans from $\sim1$ to $12~\mathrm{K}$, with $\sim80\%$ of pixels below 4~$\mathrm{K}$. 
This suggests that most of the $^{12}$CO emission is likely sub-thermally excited, which is consistent with the analysis of the Taurus MC studied by~\cite{2008ApJ...680..428G}.
In the $^{13}$CO-dark regions, $\sim 60\%$ of pixels fall within the $^{12}$CO temperature of $1.5$--$3.0\,\mathrm{K}$, with a median of $2.7\,\mathrm{K}$.
The $^{13}$CO temperature, however, is confined to a range of 0.5~$\mathrm{K}$ to 5~$\mathrm{K}$, with a peak near 1~$\mathrm{K}$.
A notable finding is that most $^{13}$CO emission arises in regions where the $^{12}$CO brightness temperature exceeds 4~$\mathrm{K}$.

\subsection{Global Distribution}\label{sec:3.2}
\begin{figure}[ht]
    \centering
    {\includegraphics[width=0.99\textwidth]{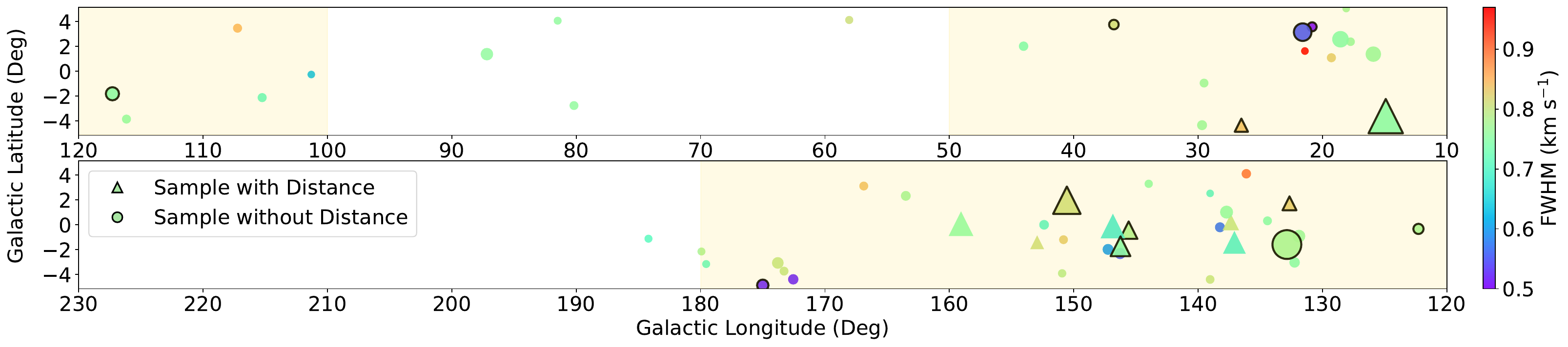}} 
    \vspace{0.2cm} 
    {\includegraphics[width=0.99\textwidth]{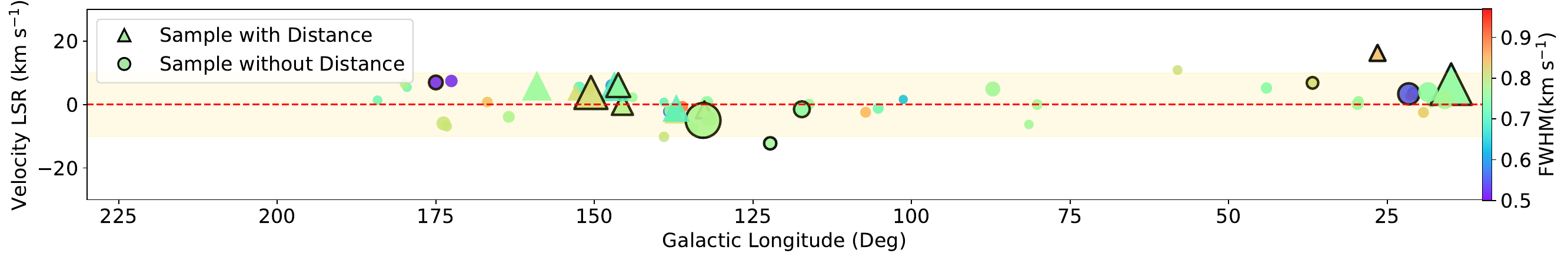}} 
    \caption{
    Global distribution of the 57 structures.
    The top panel shows their positions in Galactic longitude--latitude ($l$--$b$) space, while the bottom panel displays their distribution in longitude--velocity ($l$--$v$). 
    The symbols are the same as in Figure~\ref{fig:T-FWHM}, with the color representing the observed $^{12}$CO linewidth.
    The light-gold shaded region in the $l$--$b$ panel marks spatial concentrations ($l = 10^\circ$--$50^\circ$ and $l = 100^\circ$--$180^\circ$), while that in the $l$--$v$ panel corresponds to the velocity confinement ($|V_{\rm LSR}| \leq 10\,\mathrm{km\,s^{-1}}$).}
    \label{fig:ppv distribution}
\end{figure}

Based on the positions, sizes, $V_{\mathrm{LSR}}$, and physical distributions of our sample~(see Table~\ref{tab:1} for methodological details), we can better characterize their overall properties.
Figure~\ref{fig:ppv distribution} presents a census of 57 identified CO structures from the MWISP survey.
These structures reside primarily in two Galactic longitude intervals of $l \sim 100^\circ$--$180^\circ$ and $l \sim 10^\circ$--$50^\circ$. 
Other properties (e.g., line width, size, and temperature) show no significant variation between them.
Intriguingly, the $V_{\mathrm{LSR}}$ distribution as a function of Galactic longitude closely matches the well-known kinematic signature of the Gould Belt~\citep{2003A&A...404..519P}.

Combined with the findings of~\cite{2021MNRAS.500.3064S}, we confirm that the spatial distribution of these diffuse CO structures is highly asymmetric, concentrated in the ranges $l \approx 330^\circ$--$50^\circ$ and $l \approx 100^\circ$--$180^\circ$. 
In longitude--velocity space, the $V_{\mathrm{LSR}}$ of most of these structures is concentrated near $0\,\mathrm{km\,s^{-1}}$, predominantly within $\pm10\,\mathrm{km\,s^{-1}}$, although a few exhibit $|V_{\mathrm{LSR}}| \geq 10\,\mathrm{km\,s^{-1}}$.
By comparing the morphology of 11 larger-scale CO structures (covering nearly 50\% of the total $^{12}$CO area of our sample) with the three-dimensional extinction maps from~\cite{2025AJ....170..185Z}, we determined their distance distribution.
As shown in Figure~\ref{fig:distance}, their distances range from $196\,\mathrm{pc}$ to $265\,\mathrm{pc}$, with a typical uncertainty of approximately 10\%. 
At these distances, the MWISP data provide a spatial resolution of $\sim 0.06\,\mathrm{pc}$ and a velocity resolution of $\sim 0.16\,\mathrm{km\,s^{-1}}$.

Interestingly, we find that the global distribution of these structures is closely related to the spatial distribution and distance range of several nearby MC complexes.
For instance, in the direction of the Galactic anticenter, several structures at negative Galactic latitudes near $l \sim 170^\circ\text{--}180^\circ$ and $V_{\rm LSR} \approx 7~\mathrm{km\,s^{-1}}$ may represent a low-latitude extension of the Taurus MC complex~\citep{2008ApJ...680..428G}.
Toward the Galactic center direction, these structures are strongly concentrated in $l \sim 10^\circ\text{--}20^\circ$ and $V_{\rm LSR} \sim 0\text{--}5~\mathrm{km\,s^{-1}}$, likely suggesting an association with the Ophiuchus and Corona Australis MCs~\citep{2011ApJS..194...43P,2017ApJ...834..141O}.

For the subsample with distance measurements, the extended $^{12}$CO structures have physical sizes of approximately 1--3\,pc. 
Assuming a distance of 200\,pc, samples with smaller projected areas (i.e., $\leq$\,200\,arcmin$^2$) have radii $<$\,0.4\,pc. These sub-pc CO structures may arise as substructures at the periphery of the nearby MCs.
By cross-matching with the PGCC~\citep{2016A&A...594A..28P}, we find eight cold clumps located within the projected areas of our sample, five of which (G14.85-3.53, G14.97-3.59, G132.52-1.69, G145.52-0.46, and G158.76+0.12) are well embedded in the extended $^{12}$CO structures we studied~(ID~1, 28, 38 and 48; see Table~\ref{tab:1}).
Three of the PGCCs have angular sizes comparable to those of our sub-pc CO structures (ID 10, 37 and 54), suggesting that a minority of our samples are associated with previously known PGCCs. 
Meanwhile, more extended CO structures on parsec scales, such as that shown in Figure~\ref{fig:410845}, display diffuse $^{12}$CO emission with non-uniform brightness and hierarchical substructures.

Using the column density derived from the $X_{\mathrm{CO}}$-factor, we estimate the masses of these structures with distance measurements to range from ten to several tens of solar masses.
This is roughly comparable to the mass of the nearby cloud MBM\,40~\citep{2003NewA....8..795C,2013MNRAS.436.1152C,2025NatAs.tmp....1B}. 
However, the CO-bright MBM\,40 cloud constitutes only a small portion of the far more extensive Eos cloud, which encompasses several hundred square degrees and has a total mass of approximately $3400\,M_\odot$~\citep{2025NatAs.tmp....1B}.
Therefore, there may be a large amount of molecular gas surrounding these diffuse CO gas, and the total mass of molecular gas could significantly exceed current estimates based on CO data, possibly by one to two orders of magnitude.

\subsection{Physical Properties} \label{sec:3.3}
\subsubsection{Excitation, Optical depth, Density and Morphology
} \label{sec:3.3.1}
In this section, we use the $^{12}$CO and $^{13}$CO data to analyze the physical properties of our sample, such as their excitation conditions, optical depth distribution, and density~\citep{1991ApJ...374..540G, 1997ApJ...476..781B, 1998AJ....116..336N, 2009tra..book.....W}.

For the optically thick $^{12}$CO emission, these structures show a peak brightness temperature of $\sim7~\mathrm{K}$. This value corresponds to an estimated excitation temperature of $\sim$ 10\,K.
By assuming that $^{12}$CO and $^{13}$CO have similar excitation conditions~\citep{2010ApJ...721..686P}, the optical depth of the corresponding $^{13}$CO structures varies approximately between 0.1 and 0.4. 
The $^{12}$CO optical depth can be estimated as $\tau_{\rm ^{12}CO} \simeq X(^{12}{\rm CO}) / X(^{13}{\rm CO}) \times \tau_{\rm ^{13}CO}$, where the abundance ratio $X(^{12}{\rm CO})/X(^{13}{\rm CO})$ is adopted as 70~\citep{1994ARA&A..32..191W,2005ApJ...634.1126M}.
The derived $\tau_{\rm ^{12}CO}$ of 7--28 confirms that the $^{12}$CO emission is optically thick in the $^{13}$CO-bright regions.
The $^{12}$CO optical depth in the $^{13}$CO-dark region can be extrapolated from that at the bright edge of the $^{13}$CO emission~\citep{2023ApJ...958....7Y}. 
The $^{12}$CO line emission profiles in both regions are described by the radiative transfer equation:
\begin{equation}
T_{\mathrm{mb},\,^{12}\mathrm{CO}} = f \big[ J_\nu(T_{\mathrm{ex}}) - J_\nu(T_{\mathrm{bg}}) \big] \left(1 - e^{-\tau_{^{12}\mathrm{CO}}} \right) + J_\nu(T_{\mathrm{bg}}),
\label{eq:13dark}
\end{equation}
\begin{equation}
T_{\mathrm{mb},\,^{12}\mathrm{CO},0} = f_{0} \big[ J_\nu(T_{\mathrm{ex,0}}) - J_\nu(T_{\mathrm{bg}}) \big] \left(1 - e^{-\tau_{^{12}\mathrm{CO},0}} \right) + J_\nu(T_{\mathrm{bg}}),
\label{eq:13bright}
\end{equation}
where $J_\nu(T) = \frac{h\nu/k}{\exp(h\nu/kT) - 1}$, $T_{\rm bg} = 2.7\,\mathrm{K}$ denotes the cosmic microwave background temperature, $f$ is the filling factor, quantities with subscript $0$ refer to the edge of the $^{13}$CO-bright region, and those without subscript denote the $^{13}$CO-dark region.
Combining Equations~\ref{eq:13dark} and~\ref{eq:13bright} and defining 
$F = \frac{f\left[J_\nu(T_{\mathrm{ex}}) - J_\nu(T_{\mathrm{bg}})\right]}{f_0\left[J_\nu(T_{\mathrm{ex},0}) - J_\nu(T_{\mathrm{bg}})\right]}$,
we obtain the following relation:
\begin{equation}
\frac{T_{\mathrm{mb},\,^{12}\mathrm{CO}}-J_{\nu}(T_\mathrm{bg})}{T_{\mathrm{mb},\,^{12}\mathrm{CO},0}-J_{\nu}(T_\mathrm{bg})} 
= F \, \frac{1 - e^{-\tau_{^{12}\mathrm{CO}}}}{1 - e^{-\tau_{^{12}\mathrm{CO},0}}}.
\label{eq:tau_ratio}
\end{equation}
Pixel-by-pixel statistics reveal that the left-hand side ratio is largely confined to 0.25--0.35, while $\tau_{^{12}\mathrm{CO},0}$ on the right-hand side falls mostly between 4 and 8. 
By employing a Monte Carlo sampling combined with two-dimensional kernel density estimation (KDE) to evaluate Eq.~\ref{eq:tau_ratio} under these constraints, we find that the factor $F$ ranges from 0.25 to 0.5, corresponding to a $^{12}$CO optical depth of approximately 1 to 5 at the 50\% KDE contour level.
Moreover, the 90\% KDE contour shows that most $\tau_{^{12}\mathrm{CO}}$ values fall within 1.3--3.5, confirming that the majority of $^{12}$CO emission from the narrow-line structures is moderately optically thick.

Taking the $X_{\mathrm{CO}}$-factor to be $2.0 \times 10^{20} \mathrm{cm^{-2}\,(K\,km\,s^{-1})^{-1}}$, the $\mathrm{H_2}$ column density of our sample in the $^{13}$CO-dark regions exhibits a positively skewed distribution, with about 68\% of the pixels confined to the interval $1$--$5 \times 10^{20}\,\mathrm{cm}^{-2}$ and a peak near $3 \times 10^{20}\,\mathrm{cm}^{-2}$~\citep[$\pm 30\%$ uncertainty;][]{2013ARA&A..51..207B}.
Alternatively, the column density can be calculated from the $^{13}$CO emission under the local thermodynamic equilibrium (LTE) assumption. 
Adopting an abundance ratio of $\mathrm{H_2}$/$^{12}$CO = $1.1 \times 10^4$~\citep{1982ApJ...262..590F}, we find that, excluding a small number of very dense regions (with column densities exceeding $1 \times 10^{21}\,\mathrm{cm^{-2}}$, derived from  $^{13}$CO emission), more than 70\% of the pixels exhibit column densities in the range $1$--$5 \times 10^{20}\,\mathrm{cm}^{-2}$, with a median value of $ \sim 3 \times 10^{20}\,\mathrm{cm^{-2}}$.

Moreover, extinction measurements offer an independent check on the column density from the above CO data.
Based on the extinction map from~\cite{2025AJ....170..185Z}, we derived extinction values for the 11 CO structures with larger angular size with distance measurements~(see Section~\ref{sec:3.3}). 
For each CO structures, we estimated the mean extinction within $\pm50\,\mathrm{pc}$ of its adopted distance, yielding $\langle A_V \rangle \approx 0.73$~mag for our sample.
Using the conversion relation $N_{\mathrm{H}}\,(\mathrm{cm}^{-2}) = 2.21 \times 10^{21}\,A_V\,(\mathrm{mag})$ from \cite{2009MNRAS.400.2050G}, we derive $\langle N(\mathrm{H}) \rangle \approx 1.63 \times 10^{21}~\mathrm{cm}^{-2}$.
From $\mathrm{H\,I}$ data~\citep{2016A&A...594A.116H}, we estimate the mean $\mathrm{H\,I}$ column density within a 10~km~s$^{-1}$ velocity range for these sample to be $\langle N(\mathrm{H\,I}) \rangle \approx 8.2 \times 10^{20}~\mathrm{cm}^{-2}$, or approximately half of the total hydrogen column density inferred from extinction.
Assuming $\langle N(\mathrm{H}) \rangle = \langle N(\mathrm{H\,I}) \rangle + 2\langle N(\mathrm{H_2})\rangle$, we calculate a conservative upper limit for the average molecular hydrogen column density of $\langle N(\mathrm{H_2}) \rangle \approx 4.0 \times 10^{20}~\mathrm{cm}^{-2}$.
Taken together, we adopt a typical $\mathrm{H_2}$ column density of $\sim\!3 \times 10^{20}\,\mathrm{cm}^{-2}$ for our sample dominated by $^{12}$CO emission.

Following the method outlined by~\cite{2015PASP..127..299S}, we estimate the critical density of these structures in the $^{13}$CO-dark regions.
Under the two-level approximation and assuming a $^{12}$CO optical depth of $\tau_{^{12}\mathrm{CO}}\sim 1.3$--$3.5$ and a kinetic temperature of $10~\mathrm{K}$, the critical $\mathrm{H_2}$ density ranges from about $300$ to $1000~\mathrm{cm^{-3}}$, depending on geometry (i.e., plane-parallel slab, expanding spherical shell, and static sphere;~\citealt{2007A&A...468..627V}).
For a typical $\mathrm{H_2}$ column density of $N(\mathrm{H_2}) \approx 3 \times 10^{20}~\mathrm{cm^{-2}}$, the implied physical thickness of these structures is $\sim 0.1$--$0.3~\mathrm{pc}$.
These diffuse CO structures with large projected areas have physical extents of 1--3\,pc (Section~\ref{sec:3.2}). 
Given their sub-parsec thickness along the line of sight, these structures are characterized by a sheet-like morphology.
This configuration is naturally reproduced by the plane-parallel slab geometry, for which the predicted density is $n_{\mathrm{H_2}} \sim 300~\mathrm{cm^{-3}}$. 
We therefore adopt this value as the fiducial density for our subsequent analysis.

\subsubsection{Subsonic and Transonic Turbulence
} \label{sec:3.3.2}
The opacity significantly affects the observed spectral linewidth, particularly for optically thick line emission such as $^{12}$CO ($J=1$--$0$) \citep{1979ApJ...231..720P,2008ApJ...679..481P,2016A&A...591A.104H}.
Consequently, it is imperative to evaluate the potential impact of $^{12}$CO opacity on the observed linewidth broadening, which can be defined as~\citep{1979ApJ...231..720P,2016A&A...591A.104H}:
\begin{equation}
\beta_{\tau} = \frac{\Delta V}{\Delta V_{\mathrm{int}}} = \frac{1}{\sqrt{\ln 2}} \left[ \ln \left( \frac{\tau}{\ln \left( \frac{2}{e^{-\tau} + 1} \right)} \right) \right]^{1/2}.
\end{equation}
Here, $\Delta V$ and $\Delta V_{\mathrm{int}}$ denote the observed and the intrinsic linewidth, respectively. 

\begin{figure}[ht]
    \centering
    \includegraphics[page=1, width=0.9\textwidth]{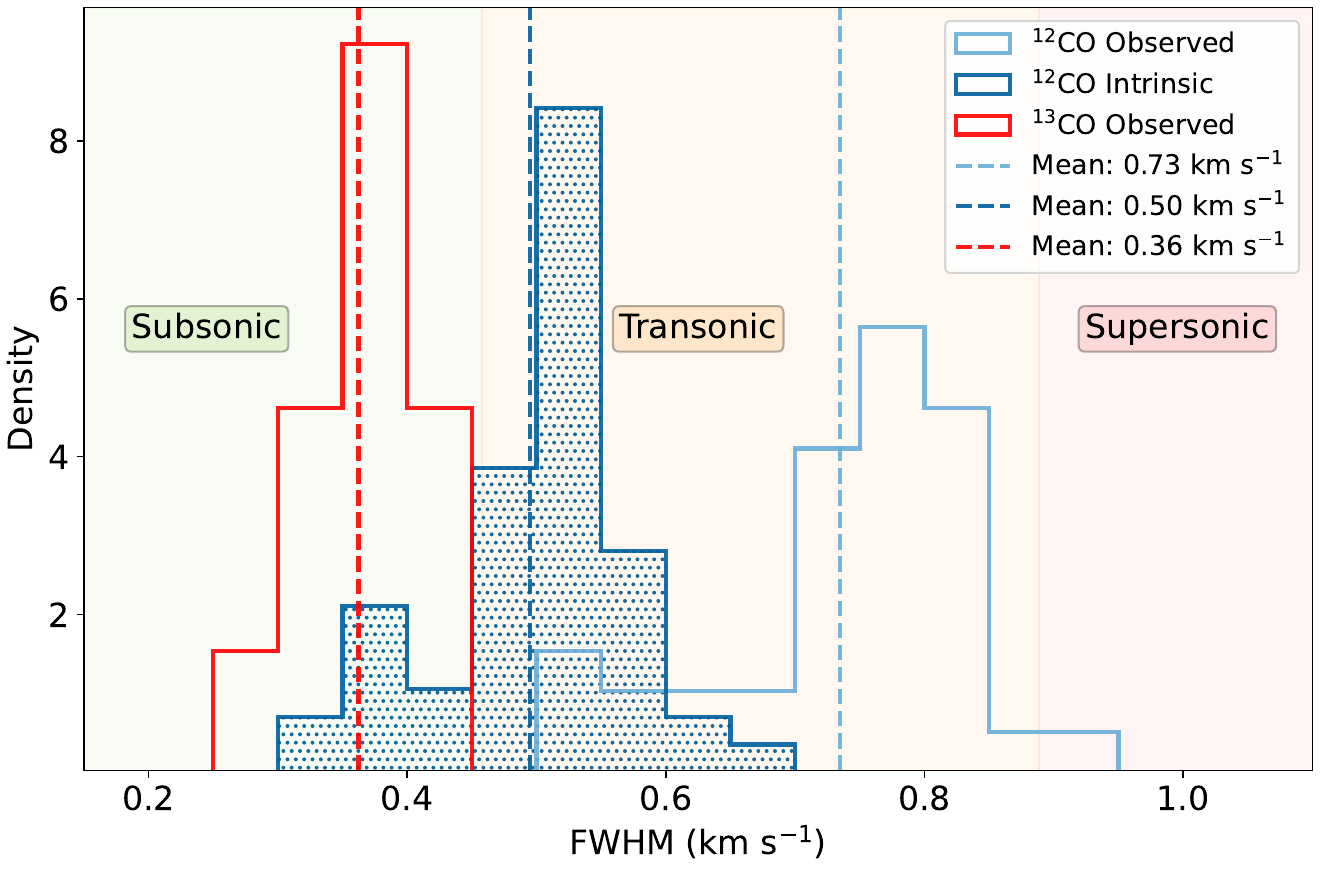}
    \caption{Distribution of CO linewidths for the 57 structures. The light and dark blue solid lines show the observed and optical-depth-corrected $^{12}$CO linewidths, respectively. The red solid line represents the observed $^{13}$CO linewidths for the 13 clouds with a projected area $>10~\mathrm{arcmin}^2$. Vertical dashed lines mark the corresponding mean values for each distribution.
    The color shadings mark the velocity intervals corresponding to subsonic, transonic, and supersonic regimes for a gas temperature of $10\,\mathrm{K}$.
    }
    \label{fig:fwhm dis}
\end{figure}

After correcting for opacity effects, the average intrinsic linewidth of $^{12}$CO in the $^{13}$CO-bright regions is $0.37\,\mathrm{km\,s^{-1}}$, consistent with the $0.36\,\mathrm{km\,s^{-1}}$ linewidth observed in optically thin $^{13}$CO.
In contrast, the intrinsic linewidth of $^{12}$CO lines in the $^{13}$CO-dark regions averages $0.52~\mathrm{km\,s^{-1}}$. 
For the 57 $^{12}$CO structures, the mean corrected linewidth is about $0.50\,\mathrm{km\,s}^{-1}$~(see Figure~\ref{fig:fwhm dis}).
It should be noted that our observations have a channel width of $0.16\,\mathrm{km\,s^{-1}}$. 
Given that the intrinsic linewidth of these diffuse CO structures is approximately two to three times the channel width, they are well spectrally resolved in our data.

Assuming a Gaussian line profile, the one-dimensional velocity dispersion~\footnote{In this work, $\sigma_{\mathrm{int}}$, $\sigma_{\mathrm{th}}$, and $\sigma_{\mathrm{NT}}$ refer to one-dimensional velocity dispersions of the gas along the line of sight rather than three-dimensional velocity dispersions.} is related to the intrinsic linewidth by $ \sigma_{\mathrm{int}} = \Delta V_{\mathrm{int}} / \sqrt{8 \ln 2} $.
This dispersion arises from the joint contribution of thermal and non-thermal motions:
\begin{equation}
\sigma_{\mathrm{int}} = \sqrt{\sigma_{\mathrm{th}}^2 + \sigma_{\mathrm{NT}}^2}.
\end{equation}
Here, $\sigma_{\mathrm{th}} = \sqrt{k T / (\mu_{\mathrm{CO}} m_{\mathrm{H}})}$ is the thermal velocity dispersion, where $k$ is the Boltzmann constant, $T$ is the gas kinetic temperature, $\mu_{\mathrm{CO}} = 28$ for CO, and $m_{\mathrm{H}}$ is the mass of a hydrogen atom. 
Assuming a gas kinetic temperature of $T = 10\,\mathrm{K}$, the thermal velocity dispersion is  $ \sigma_{\mathrm{th}} \approx 0.05\,\mathrm{km\,s^{-1}} $ , and the isothermal sound speed is  $ c_{\mathrm{s}} = \sqrt{\gamma k T / (\mu m_{\mathrm{H}})} \approx 0.19\,\mathrm{km\,s^{-1}} $ for $ \mu = 2.37 $~\citep{2008A&A...487..993K} and $ \gamma = 1 $~\citep{2005ism..book.....L}.
The magnitude of non-thermal motions can be measured by the Mach number, defined as $ \mathcal{M} = \sigma_{\mathrm{NT}} / c_{\mathrm{s}} $.
The calculations show that in the $^{13}$CO-bright regions, these narrow-line CO structures have $ \mathcal{M} \lesssim 1 $ , indicating subsonic motion. 
The average Mach number for all identified $^{12}$CO structures is close to 1, indicating a transition from subsonic to transonic turbulence~(see Figure~\ref{fig:fwhm dis}).

Even including velocity gradients, the velocity dispersion of the CO structures in our sample remains only about $0.2$--$0.4\,\mathrm{km\,s^{-1}}$ ($M \sim 1$--$2$, assuming $T \sim 10\,\mathrm{K}$; Column~(5) in Table~\ref{tab:1}).
This value is notably lower than the supersonic turbulence commonly observed in MCs, suggesting that the molecular gas is relatively quiescent compared to the general MC population. 
On parsec scales, the extended, diffuse structures exhibit transonic to mildly supersonic turbulence, placing them in a low-velocity-dispersion regime (see Section~\ref{sec:4.2} for further discussion).
We refer to these diffuse, sheet-like structures as veil clouds, which, together with the smaller $^{12}$CO structures discussed below, are a primary focus of our analysis.

\section{discussion}\label{sec:discussion}
\subsection{Association with the Local Bubble} \label{sec:4.1}
Recent studies have revealed that a series of nearby MC complexes are situated on the surface of the Local Bubble~\citep{2022Natur.601..334Z,2024ApJ...973..136O}.
These MCs, including those in Ophiuchus, Lupus, Pipe, Chamaeleon, Corona Australis, and Taurus, all lie within approximately 200\,pc~\citep{2021ApJ...919...35Z}.
The CO structures with narrow lines, predominantly veil clouds, in this work show a spatial concentration along sightlines toward the Galactic center and anticenter.
This distribution aligns with both the local extinction map within $\sim 300\,\mathrm{pc}$~\citep{2025AJ....170..185Z} and the positions of the aforementioned nearby MC complexes.
This correlation extends beyond our sample to include other subsonic MCs, such as MBM 40 within the Eos cloud~\citep{2025NatAs.tmp....1B} and the Musca cloud, the latter being part of a larger C-shaped structure~\citep{2024A&A...687L...9E}. 
Critically, all these nearby MCs are situated close to the Local Bubble. 
Hence, we conclude that the majority of these sheet-like,  diffuse CO structures are located within $300\,\mathrm{pc}$ (and likely closer).

\begin{figure}[ht]
    \centering
    \includegraphics[page=1, width=0.9\textwidth]{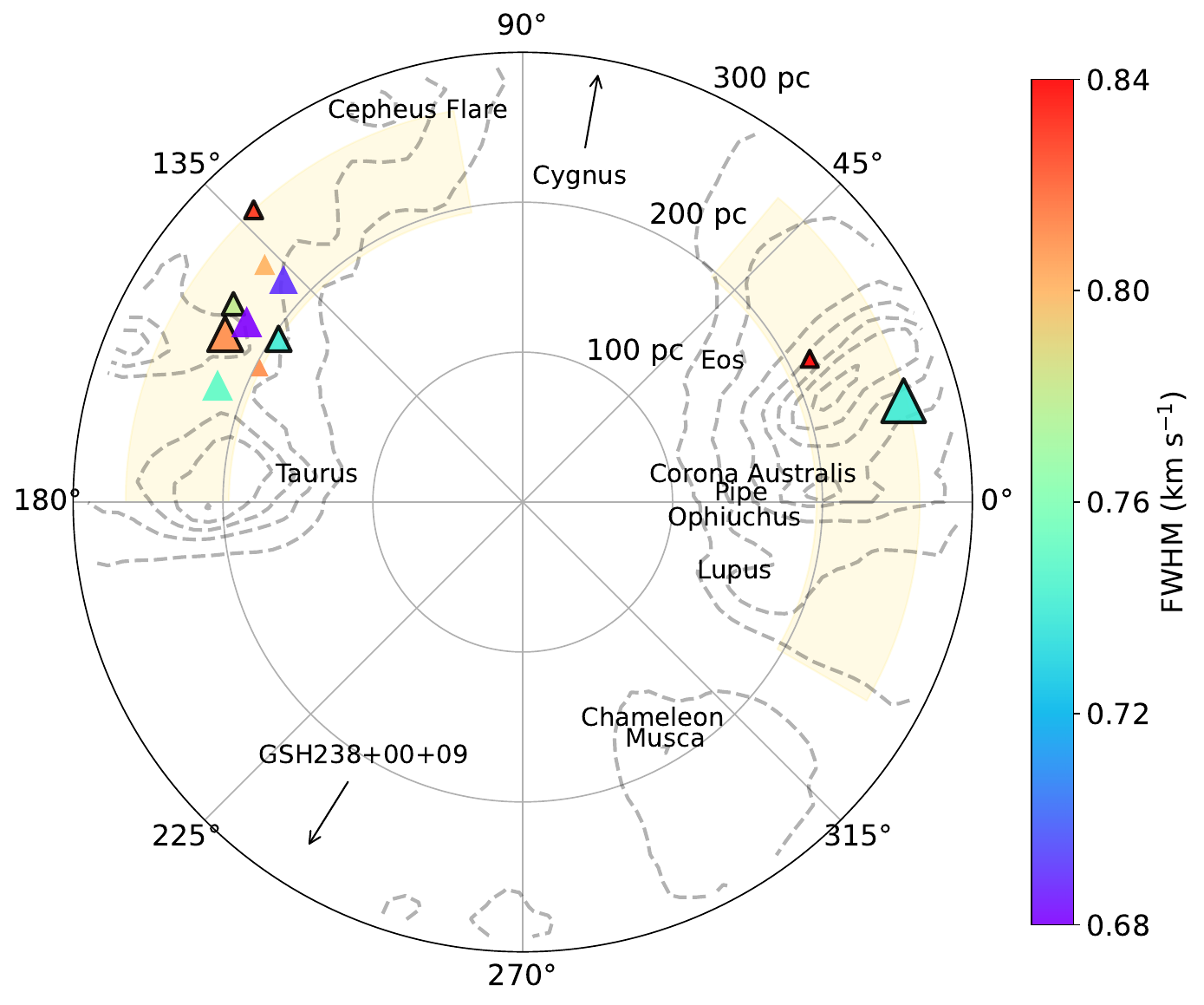}
    \caption{
    Face-on view of the 11 clouds with distance measurements. 
    The symbols are the same as in Figure~\ref{fig:ppv distribution}. 
    The light-gold shaded regions mark the overall distribution of these clouds, spanning $196$--$265~\mathrm{pc}$.
    Gray dashed contours represent extinction structures within $|b| < 30^\circ$, starting from $0.3~\mathrm{mag}$ with an interval of $0.3~\mathrm{mag}$~\citep{2025AJ....170..185Z}. 
    Marked are the positions of nearby MCs (e.g., Taurus, Ophiuchus, and Chameleon; also see the Cepheus Flare in \citealt{2025AJ....170...24W} and \citealt{2025AJ....170..185Z}), as well as the directions toward the radio supershell GSH\,$238+00+09$~\citep{1998ApJ...498..689H} and the Cygnus region~\citep{2024AJ....167..220Z}.
    }
    \label{fig:distance}
\end{figure}

\begin{figure}[ht]
    \centering
    \includegraphics[width=0.9\textwidth]{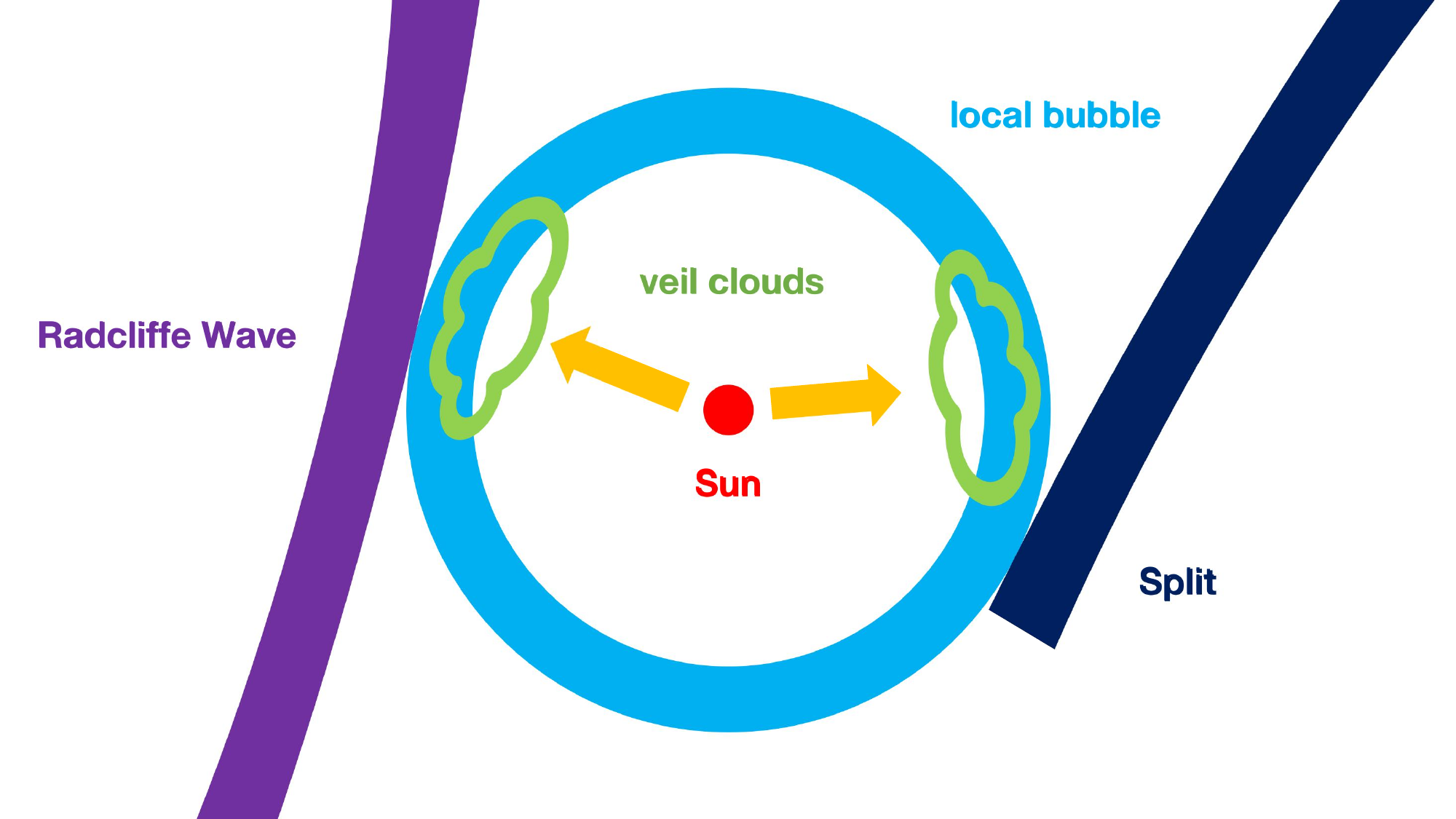} 
    \caption{
    Schematic of veil clouds relative to the Local Bubble and other large-scale structures.
    The red circle shows the Solar System, while orange arrows indicate two sightlines rich in such clouds.
    The boundary of the Local Bubble is outlined by the blue circle. 
    Purple and dark blue strips represent the Radcliffe Wave and the Split, respectively.
    Green cloud-shaped symbols denote regions where veil clouds are densely clustered adjacent to the Local Bubble.
    }
    \label{fig:local bubble}
\end{figure}

These narrow-line structures exhibit a coherent spatial arrangement that closely follows the local gas distribution traced by dust extinction (Figure~\ref{fig:distance}). 
This alignment is consistent with these structures outlining the periphery of the Local Bubble. 
These nearby diffuse MCs may thus be associated with superbubbles driven by recent star formation activity, such as the clouds in the Taurus--Perseus complex toward the Galactic anticenter and the Scorpius--Centaurus complex toward the Galactic center~\citep{2016Natur.532...69W,2021ApJ...919L...5B,unknown,2024Natur.631...49S}.
They are likely components of larger Galactic features such as the Radcliffe Wave and the Split~(\citealt{2020Natur.578..237A}, see also Figure~\ref{fig:local bubble}).
In contrast, between the Radcliffe Wave and the Split, roughly at $l \sim 70^\circ$ and $240^\circ$, our samples are sparse (see Figure~\ref{fig:distance}).
This region is traversed by chimneys and tunnels connecting the Local Bubble to neighboring cavities (e.g., GSH\,238+00+09; \citealt{2003A&A...411..447L,2015JPhCS.577a2016L,2024ApJ...973..136O}),
where the low-density interstellar gas naturally explains the scarcity of structures with narrow lines.

We propose that the formation of these sheet-like structures is most likely tied to episodic supernova explosions within the Local Bubble~\citep{2002PhRvL..88h1101B,2006MNRAS.373..993F}. 
Specifically, the resulting shocks compress the warm neutral medium, creating conditions that favor thermal instability and the subsequent formation of cold neutral medium~\citep{2000ApJ...532..980K, 2002ApJ...564L..97K, 2009ApJ...695..825I}.
This is followed by rapid radiative cooling and a phase transition, giving rise to the cold, dense sheet-like structures~\citep{2006ApJ...643..245V} at the periphery of the Local Bubble.
Within these compressed sheets, the enhanced density, coupled with increased self-shielding and dust shielding, facilitates the formation of molecular gas~\citep{1996ApJ...468..269D,2010ApJ...716.1191W}.
It is worth noting that we detected $\mathrm{H\,I}$ narrow self-absorption~(HINSA;~\citealt{2003ApJ...585..823L, 2005ApJ...622..938G}) features toward several of our samples.
This suggests that these diffuse CO structures trace the transition from atomic to molecular gas and represent an early stage of MC formation.
However, a systematic analysis of HINSA for the entire sample is beyond the scope of this work. 
We leave this for future investigation.

\subsection{Quiescent Regions and Turbulent Dissipation}\label{sec:4.2}
The low velocity dispersions of our sample suggest that these regions are quiescent, where sustained turbulent injection is likely absent.
As a result, the supersonic turbulence at the cloud scales cascades to smaller scales, where the velocity dispersion is lower~\citep{1981MNRAS.194..809L}.
Adopting a progenitor structure with a typical scale of $1\,\mathrm{pc}$ and the corresponding velocity dispersion from~\cite{1981MNRAS.194..809L}, we estimate a turbulent cascade crossing timescale of $t_{\rm cross} \approx L / \sigma_v \lesssim 1\,\mathrm{Myr}$.
Because supernova explosions in the Local Bubble are episodic, with the most recent major episode occurring roughly $2.5\,\mathrm{Myr}$ ago~\citep{2016Natur.532...73B,2016PNAS..113.9232L,2020PhRvL.125c1101K}, any turbulence injected by past activity has long since decayed.
Thus, at the peripheries of their parent MCs, supersonic turbulence can cascade from large to small scales, ultimately fragmenting into compact, dynamically quiescent structures (possibly analogous to cloud ID 54 associated with PGCC G175.00-4.83; and the CO structure may lie at a distance of $235 \pm 33$\,pc;~\citealt{2026ApJS..284...71M}).
In contrast, molecular gas in the main regions of cloud complexes may experience sustained turbulent injection and mass accumulation (e.g., through cloud--cloud collisions, gravitational instabilities, and star formation feedback, etc.), which continuously replenish turbulent energy and maintain supersonic turbulence.

For veil clouds, the velocity dispersion of the pc-scale CO structures is generally less than twice the sound speed at about 10\,K (Table~\ref{tab:1}), substantially lower than that of typical MCs of comparable size~\citep{1981MNRAS.194..809L}.
If magnetic fields are taken into account, ion--neutral friction may provide an additional efficient mechanism for turbulent dissipation, particularly on the sub-parsec scales~\citep{2019FrASS...6....5H}.
The $B$--$n$ relation~\citep{2010ApJ...725..466C,2012ARA&A..50...29C} indicates that in diffuse gas (e.g., for veil clouds with typical densities of $n_{\mathrm{H_2}} \sim 300\,\mathrm{cm^{-3}}$ here) the magnetic field strength does not scale with density and typically remains on the order of several to roughly $10\,\mu\mathrm{G}$.
The typical magnetic field strength in the Local Bubble, on whose surface the veil clouds reside, is $\sim 6~\mu$G~\citep{2006ApJ...640L..51A}.
Meanwhile, nearby MCs on the surface of the Local Bubble traced by $^{12}$CO exhibit stronger fields: $\sim 14\,\mu\mathrm{G}$~\citep{2008ApJ...680..420H} in the diffuse envelope of Taurus and $\sim 12\,\mu\mathrm{G}$ in MBM\,40 within Eos~\citep{2025ApJ...992..132K}.
Interestingly, some works report magnetic field strengths of 50--70~$\mu\mathrm{G}$ in the nearby Riegel--Crutcher cloud~\citep{2006ApJ...652.1339M, 2026ApJ..1004...70F}, a cold neutral medium layer located toward the Galactic center.
We have also identified associated sheet-like CO structures toward the cold $\mathrm{H\,I}$ cloud (in prep.), further underscoring the dynamical role of magnetic fields in diffuse gas. 
We therefore adopt a magnetic field strength of $\gtrsim 10\,\mu\mathrm{G}$ as a reasonable working assumption for veil clouds.

For these diffuse clouds with $T \sim 10~\mathrm{K}$ and $N(\mathrm{H_2}) \sim 3 \times 10^{20}\,\mathrm{cm^{-2}}$, the magnetic pressure $P_B \sim 3 \times 10^4 \, \mathrm{K\,cm^{-3}}$ exceeds the thermal pressure $P_{\mathrm{th}} \sim 3 \times 10^3 \, \mathrm{K\,cm^{-3}}$, which is in good agreement with the average thermal pressure of $3070 \, \mathrm{K\,cm^{-3}}$ at the solar circle \citep{2003ApJ...587..278W}, by more than an order of magnitude, and the mass-to-flux ratio is $\sim 0.6$.
This suggests that magnetic fields may dominate the dynamical state of veil clouds.
Following~\cite{2013A&A...556A.153H}, we can estimate the dissipation scale:
\begin{equation}
l_{\mathrm{diss}} = \left( \frac{L_0^{1/3}}{\rho_0^{1/3} V_0} \right)^{3/4} \left( \frac{B^2}{4\pi \gamma_{\mathrm{ad}} \rho^{2/3} \rho_i} \right)^{3/4},
\label{eq:l_diss}
\end{equation} 
and the timescale is given by~\citep{2013A&A...560A..68H}:
\begin{equation}
\tau_{\mathrm{diss,amb}} = \frac{2\gamma_{\mathrm{ad}}\rho_i}{v_A^2 (2\pi/\lambda)^2}.
\end{equation}
Here, $L_0$ is set to a parsec scale, $V_0 = 0.9 \, L_0^{0.5}$~\citep{2003ApJ...584..293B, 2004ApJ...615L..45H}, and $\rho_0 = \mu n_0 m_{\mathrm{H}}$, where $n_0$ is of the order of $10^2\,\mathrm{cm^{-3}}$ and $\mu = 2.37$~\citep{2008A&A...487..993K}.
For veil clouds, the neutral mass density is $\rho = \mu n m_{\mathrm{H}}$ with a typical number density of $n \sim 300\,\mathrm{cm^{-3}}$ and the ion mass density is given by $\rho_i = \chi_i \mu_i n m_{\mathrm{H}}$, where we adopt $\bar{\mu}_i \sim 20$ to represent the mean molecular weight for ions such as HCO$^+$, C$^+$, and H$_3^+$.
The ion--neutral coupling is characterized by a drag coefficient of $\gamma_{\mathrm{ad}} = 3.5 \times 10^{13}\,\mathrm{cm^3\,g^{-1}\,s^{-1}}$~\citep{1981ApJ...246...48M}.
The Alfv\'en speed is defined as $ v_A = B/\sqrt{4\pi\rho} $, and the relevant scale $\lambda$ is assumed to be comparable to the characteristic size of these sheet-like MCs.
We find that the dissipation scale in these sheet-like clouds is $l_{\rm diss} \lesssim 0.2~\mathrm{pc}$, comparable to their thickness.  
The derived dissipation timescale is $\lesssim 1\,\mathrm{Myr}$, suggesting that ion–neutral friction may efficiently dissipate turbulence in these veil clouds from pc to sub-pc scales,  i.e., down to their characteristic thickness scale of 0.1--0.3\,pc.
We note that the magnetic field strength inferred here for veil clouds is based on reasonable assumptions rather than direct observational constraints. 
Further observations would be valuable to investigate the magnetic properties of this cloud population.

\subsection{Are Veil Clouds Widespread?}
These diffuse, sheet-like clouds exhibit subsonic to transonic turbulence, and may have formed through turbulent compression driven by past supernova explosions, possibly tracing an early transition from atomic to molecular gas.
Magnetic fields are dynamically important for these diffuse atomic or molecular gas structures~\citep{2023MNRAS.525..721G}.
Diffuse gas within these sheet-like structures is assembled by flows along magnetic field lines, leading to the formation of elongated, feather-like substructures whose long axes are inherently aligned with the magnetic field~\citep{2010ApJ...725..466C,2013ApJ...774..128S,2017A&A...607A...2S,2019ApJ...878..110F,2023MNRAS.525..721G}.
In general, such diffuse clouds are likely expected to disperse via photodissociation rather than undergo gravitational collapse, as studied in clouds such as the Eos cloud~\citep{2025NatAs.tmp....1B}.
Conversely, sustained energy injection can maintain and amplify turbulence within these clouds, driving its further evolution.
An ordered magnetic field renders turbulence anisotropic, channeling density fluctuations and gas flows along the field lines~\citep{2013MNRAS.436.3707L,2014prpl.conf..101L}, as observed in the magnetically aligned striations~\citep{2008ApJ...680..428G,2010A&A...518L.102A,2016A&A...590A.110C,2016MNRAS.462.1517P,2024AJ....167..176S}.
This localized mass accumulation can, in turn, enhance the role of gravity within the filamentary structures.
These filaments subsequently accrete mass and fragment into subsonic, dense cores, which are likely the progenitors of future stars~\citep{2011A&A...533A..34H,2020ApJ...896..110L,2022A&A...663A..82G}.

Our study shows that these veil clouds are spatially associated with both nearby MC complexes and, most likely, the Local Bubble. 
Supershells driven by stellar feedback are widespread structures, found not only in the Milky Way but also in external galaxies~\citep{2012MNRAS.424.2442S,2023ApJ...944L..22B}.
This ubiquity suggests that the physical conditions governing the formation of these diffuse, subsonic/transonic MCs are not confined to the solar neighborhood.
In quiescent regions, such as those outside the Central Molecular Zone and active star-forming sites, turbulence decays rapidly in the absence of sustained energy injection~(see Section~\ref{sec:4.1}).
We therefore propose that such diffuse, subsonic/transonic structures are not merely a local phenomenon, but are widespread throughout the Galaxy.

It is important to note that the diffuse interstellar medium offers only weak shielding against the ambient interstellar radiation field.
Veil clouds mark localized regions of enhanced $\mathrm{H_2}$ within this diffuse medium, where the gas density reaches the threshold necessary to shield CO from photodissociation.
Consequently, the surroundings of these diffuse CO clouds likely host substantial reservoirs of CO-dark molecular gas.
A case in point is MBM 40 in the Eos cloud~\citep{2025NatAs.tmp....1B}, which represents only the tip of the iceberg of the extensive diffuse molecular gas in which it is embedded.
If veil clouds are indeed widespread throughout the Galaxy, the total mass of molecular gas in the extended regions they inhabit may significantly exceed estimates based solely on CO-bright emission~\citep{2005Sci...307.1292G,2010ApJ...710..133A,2011A&A...536A..19P,2014MNRAS.441.1628S}.
The concentration of these diffuse MCs, distinguished by our veil cloud samples, thus reveals substantial molecular gas reservoirs, with important implications for the distribution of Galactic gas and the initial conditions of star formation.

However, veil clouds have been largely overlooked in previous studies due to observational limitations, including their intrinsically low surface brightness, insufficient spatial and spectral resolution, and limited sky coverage. 
Coarser spatial resolution probes a larger physical volume, blending the emission from multiple unresolved subsonic/transonic gas components into a single, broader spectral line. 
Meanwhile, insufficient spectral resolution hampers the separation of gas components in kinematically crowded regions. 
The apparent line broadening is therefore likely a consequence of line-of-sight volume integration and the superposition of independent velocity components, rather than solely an intrinsic increase in turbulent gas motions. 
This effect is particularly severe toward the inner Galaxy, where the higher crowding of MCs along the line of sight leads to substantial blending of distinct velocity features. 
A systematic identification of such narrow-line CO structures will require the application of more advanced analysis algorithms.

\section{summary}\label{sec:summary}
Based on the CO data from Phase I of the MWISP survey, we have systematically studied MCs with exceptionally narrow lines in the northern Galactic plane.
Our main conclusions are as follows:

\begin{enumerate}
\item 
We identified a total of 57 narrow-line CO structures with areas exceeding $56.25\,\mathrm{arcmin^2}$. 
They exhibit low turbulent velocities on sub-pc to pc scales, corresponding to subsonic/transonic to mildly supersonic turbulence.
Most of these structures are diffuse, low-density MCs, with sub-thermally excited CO emission, moderate $^{12}$CO optical depths ($\tau_{^{12}\mathrm{CO}} \sim 1.3$--$3.5$), and low H$_2$ column densities ($N_{\mathrm{H}_2} \sim 3 \times 10^{20}$~cm$^{-2}$). 
Adopting the gas volume density estimated in our study, $n_{\mathrm{H}_2} \sim 300$~cm$^{-3}$, we derive thicknesses of 0.1--0.3~pc for these structures.

\item 
These structures lie within $300\,\mathrm{pc}$ (and are likely closer), showing an asymmetric concentration toward both the Galactic center and anticenter directions. 
They are spatially associated with nearby MC complexes such as Taurus, Ophiuchus, and Corona Australis, and their velocity pattern is similar to that of the Gould Belt.
They may represent quiescent regions where, without sustained turbulent energy injection, turbulence undergoes a continuous cascade to smaller scales, leading to their observed subsonic/transonic state.

\item 
Veil clouds are diffuse, extended, sheet-like CO structures with physical sizes of $\sim$1--3~pc and low volume and column densities. 
They constitute a subsample in our dataset that is more extended than the smaller CO structures identified in this study. 
Several PGCCs are embedded as substructures within these extended sheet-like clouds, forming part of their hierarchical architecture.

\end{enumerate}

Considering that veil clouds are highly diffuse, they may provide only weak shielding against the interstellar radiation field, resulting in a relatively high ionization fraction ($\chi_i \sim 10^{-6}$ or higher; \citealt{2007ARA&A..45..339B}).
This enhances the coupling between neutrals and the magnetic field via ion--neutral collisions. 
Accordingly, on sub-parsec scales comparable to the typical thickness of veil clouds, turbulence in these clouds is likely dissipated primarily by ion--neutral friction rather than by viscous or Ohmic processes~\citep{2023MNRAS.525..721G,2024MNRAS.531.4138B}.
At scales below the ambipolar diffusion scale, ion–neutral friction suppresses turbulent fluctuations, leading to a smoother structure~\citep{2016A&A...589A..24N}.
Ion--neutral friction can drive intermittent dissipation of turbulent energy in diffuse, magnetized gas~\citep{2008ApJ...684..380L,2010MNRAS.406.1201T,2012ApJ...744...73L,2014MNRAS.443...86M}.
This process gives rise to sheet-like or ribbon-like structures consistent with the observed diffuse clouds~\citep{2004JFM...513..111M,2016ApJ...831...46A} and may help preserve their large-scale morphology by suppressing small-scale chaotic structure.
It also can heat diffuse gas locally, enhancing endothermic reactions and explaining elevated CH$^+$ and SH$^+$ abundances~\citep{2016ARA&A..54..181G}.
Future studies of the magnetic field, ionization properties, gas structures/kinematics, and molecular chemistry of veil clouds will be essential for testing the scenario outlined in this study and for understanding the physical processes governing these diffuse structures.

Veil clouds may trace the transition zone from atomic to molecular gas, representing an early evolutionary phase of MCs.
They are young MCs formed by turbulent compression driven by supernova explosions and simultaneously serve as the raw material for potential future star formation.
Magnetic fields likely dominate the evolution of these diffuse, subsonic/transonic MCs, with turbulence stretching the gas along field lines to form aligned, extended structures. 
If these clouds accumulate sufficient mass, gravity may eventually overcome magnetic support. 
This would allow initially sheet-like structures to evolve into dense filaments, establishing physical conditions conducive to subsequent star formation. 

Given the prevalence of quiescent environments throughout the Galaxy, veil clouds are likely a widespread phenomenon. 
They trace localized $\mathrm{H_2}$-enhanced pockets within the diffuse interstellar medium, where the column density reaches the threshold for shielding CO from photodissociation. 
Embedded in a vast reservoir of CO-dark gas, these structures represent merely the visible tip of a far more extensive molecular component. 
Veil clouds therefore reveal a widespread and previously overlooked population of diffuse molecular structures, with direct implications for the physics of the atomic-to-molecular transition, the total reservoir of Galactic molecular gas, and the initial conditions for star formation.

\begin{acknowledgments}
We would like to thank the anonymous referee for the valuable comments and suggestions that have largely improved the manuscript.
This research made use of the data from the Milky Way Imaging Scroll Painting (MWISP) project, which is a multiline survey in $^{12}$CO/$^{13}$CO/C$^{18}$O along the northern Galactic plane with the PMO $13.7\,\mathrm{m}$ telescope. 
We are grateful to all the members of the MWISP working group, particularly the staff members at the PMO $13.7\,\mathrm{m}$ telescope, for their long-term support. 
MWISP was sponsored by the National Key R\&D Program of China with grants 2023YFA1608000, 2017YFA0402701, and the CAS Key Research Program of Frontier Sciences with grant QYZDJ-SSW-SLH047. 
Y.S. acknowledges funding from the Basic Research Program of Jiangsu (BK20252109).
\end{acknowledgments}

\facilities{PMO:DLH}

\software{astropy~\citep{2013A&A...558A..33A,2018AJ....156..123A,2022ApJ...935..167A}}


\centerwidetable
\begin{deluxetable}{cccccccccccccc}
\tablecaption{Catalog of $^{12}$CO Structure with Narrow Lines}
\label{tab:1}
\tabletypesize{\small}
\tablehead{
\colhead{ID} &
\colhead{$l_{\rm cen}$} &
\colhead{$b_{\rm cen}$} &
\colhead{$v_{\rm cen}$} &
\colhead{$\sigma_{v}$} &
\colhead{FWHM} &
\colhead{$T_{\rm mean}$} &
\colhead{$T_{\rm peak}$} &
\colhead{$N_{\rm mean}$} &
\colhead{$Area_{\rm ang}$} &
\colhead{$Area_{\rm ang}$($^{13}$CO)} &
\colhead{Dist} &
\colhead{$R_{\rm eff}$} &
\colhead{Mass}\\
\colhead{} &
\colhead{(deg)} &
\colhead{(deg)} &
\colhead{(km\,s$^{-1}$)} &
\colhead{(km\,s$^{-1}$)} &
\colhead{(km\,s$^{-1}$)} &
\colhead{(K)} &
\colhead{(K)} &
\colhead{(cm$^{-2}$)} &
\colhead{(arcmin$^2$)} &
\colhead{(arcmin$^2$)} &
\colhead{(pc)} &
\colhead{(pc)} &
\colhead{($M_\odot$)}\\
\colhead{(1)} &
\colhead{(2)} &
\colhead{(3)} &
\colhead{(4)} &
\colhead{(5)} &
\colhead{(6)} &
\colhead{(7)} &
\colhead{(8)} &
\colhead{(9)} &
\colhead{(10)} &
\colhead{(11)} &
\colhead{(12)} &
\colhead{(13)} &
\colhead{(14)}
}
\startdata
$1^{*}_{\rm }$ & 14.91 & -3.61 & 6.48 & 0.53 & 0.74 & 4.3 & 11.8 & 6.6e+20 & 263 & 1148.25 & 301.25 & 1.46 & 99 \\
2 & 15.91 & 1.38 & 1.51 & 0.35 & 0.76 & 3.5 & 8.8 & 5.4e+20 & 228.75 & 8 &  &  &  \\
3 & 17.74 & 2.38 & 1.29 & 0.38 & 0.76 & 2.9 & 5.4 & 4.4e+20 & 72.25 &  &  &  &  \\
\enddata
\tablenotetext{}{}
\textbf{Note.} 
Column (1): ID of $^{12}$CO structures, sorted by longitude. $^*$ indicates the PGCC lies within the projected $^{12}$CO emission.
Columns (2)--(4): central coordinates in $l$--$b$--$v$ space, obtained via intensity-weighted averaging: $x_{\mathrm{cen}} = \frac{\sum_i W_i\, x_i}{\sum_i W_i}$, where the sum is over all Gaussian components indexed by $i$, $W_i$ is the integrated intensity of the $i$-th Gaussian component, and $x$ represents $l$, $b$, or $v$.
Columns (5): overall velocity dispersion of the $^{12}$CO structure, defined as $\sigma_{v}^{2} = \frac{\sum_{j} T_{j} (v_{j} - v_{\mathrm{cen}})^{2}}{\sum_{j} T_{j}}$, where $T_j$ and $v_j$ denote the temperature and velocity of the $j$-th voxel in the PPV space, respectively.
Columns (6): intensity-weighted FWHM computed pixel by pixel, defined as $\mathrm{FWHM} = 2\sqrt{2\ln 2} \, \frac{\sum_i W_i \, \sigma_i}{\sum_i W_i}$, where $\sigma_i$ is the standard deviation of the $i$-th Gaussian component . 
Columns (7)--(8): the mean and maximum values of $T_{\rm mb}$ obtained from a pixel-by-pixel analysis across all detected emission regions.
Columns (9):the mean H$_2$ column density estimated using the $X_{\mathrm{CO}} = 2 \times 10^{20}~\mathrm{cm^{-2}\,(K\,km\,s^{-1})^{-1}}$.
Column (10)--(11): areas of both $^{12}$CO structures and the enclosed $^{13}$CO structures. 
Columns (12): distance of $^{12}$CO structures, with an uncertainty of approximately 10\%.
Columns (13): physical radius of $^{12}$CO structures.
Columns (14): mass calculated by $^{12}$CO emission with determined distance.

\end{deluxetable}

\bibliography{sample701}{}
\bibliographystyle{aasjournalv7}

\end{document}